\documentclass[conference]{IEEEtran}
\IEEEoverridecommandlockouts
\usepackage{cite}
\usepackage{amsmath,amssymb,amsfonts}
\usepackage{algorithmic}
\usepackage{graphicx}
\usepackage{textcomp}
\usepackage{xcolor}
\usepackage{siunitx}
\graphicspath{{..figures/}{../figures/}}
\def\BibTeX{{\rm B\kern-.05em{\sc i\kern-.025em b}\kern-.08em
    T\kern-.1667em\lower.7ex\hbox{E}\kern-.125emX}}
\begin{document}

\title{Towards In-Air Ultrasonic QR Codes:\\ 
Deep Learning for Classification of\\ Passive Reflector Constellations
}

\author{\IEEEauthorblockN{Wouter Jansen}
\IEEEauthorblockA{\textit{FTI Cosys-Lab, University of Antwerp}\\ Antwerp, Belgium \\
\textit{Flanders Make Strategic Research Centre}\\ Lommel, Belgium\\
wouter.jansen@uantwerpen.be}
\and
\IEEEauthorblockN{Jan Steckel}
\IEEEauthorblockA{\textit{FTI Cosys-Lab, University of Antwerp}\\ Antwerp, Belgium \\
\textit{Flanders Make Strategic Research Centre}\\ Lommel, Belgium\\
jan.steckel@uantwerpen.be}
}

\maketitle

\begin{abstract}
In environments where visual sensors falter, in-air sonar provides a reliable alternative for autonomous systems. While previous research has successfully classified individual acoustic landmarks, this paper takes a step towards increasing information capacity by introducing reflector constellations as encoded tags. Our primary contribution is a multi-label Convolutional Neural Network (CNN) designed to simultaneously identify multiple, closely spaced reflectors from a single in-air 3D sonar measurement. Our initial findings on a small dataset confirm the feasibility of this approach, validating the ability to decode these complex acoustic patterns. Secondly, we investigated using adaptive beamforming with null-steering to isolate individual reflectors for single-label classification. Finally, we discuss the experimental results and limitations, offering key insights and future directions for developing acoustic landmark systems with significantly increased information entropy and their accurate and robust detection and classification.
\end{abstract}


\section{Introduction}\label{sec:intro}
Optical sensors, like cameras, can establish a priori ground truth for localization and mapping by using information-encoded tags such as QR codes. These tags, also called landmarks, can significantly improve the results of these applications \cite{thrun_probabilistic_2005} and can correct the accumulated error in the mapping process \cite{grisetti_tutorial_2010}. Light-based sensors, however, struggle in adverse environmental conditions like rain, fog, or dust, whereas in-air ultrasonic sensing offers a robust alternative \cite{yoneda_automated_2019, zhang_perception_2023, everett_sensors_1995, jansen_real-time_2022}. While modern 3D in-air ultrasonic sensors can create high-resolution scans of the environment using dense microphone arrays \cite{laurijssen_hiris_2024, allevato_air-coupled_2022, kerstens_ertis_2019}, recognizing the environment's semantics remains a major challenge. The echo's characteristics are highly dependent on an object's material, structure, and orientation\cite{eisele_relevance_2024, werner_g_neubauer_acoustic_1986, simmons_acoustic_1989}, which make multi-object classification with in-air ultrasound sensors a complex challenge. As such, passive in-air ultrasound tags with high information entropy, such as QR-codes for optical sensors, and the algorithms to recognize them in real-time, do, to the best of our knowledge, not exist yet for in-air ultrasonic sensors. 

Previous work has explored deep learning for acoustic object classification, often using objects with easily differentiable structures \cite{dror_three-dimensional_1995, kroh_classification_2019, ayrulu_neural_2001}. Our own prior research focused on classifying various sizes of a single object structure which was bio-inspired by a dish-shaped leaf with a conspicuous acoustic echo of the Cuban liana \textit{Marcgravia evenia}, a bat-pollinated plant \cite{simon_bioinspired_2020}. These leaves were approximated using 3D-printed shapes. Several classification methods were explored to classify these landmarks and their radius. In a first attempt, a support vector machine was used for real-time classification and detection with four different radius variations, reaching results of around 67.2\% accuracy \cite{de_backer_detecting_2023}. This effort culminated in the creation of a single-label Convolutional Neural Network (CNN) that achieved 97.3\% classification accuracy for ten different reflector sizes and predicted their orientation with an RMSE below \SI{10}{\degree} using a regression output \cite{jansen_semantic_2024}. We used a cochleogram representation in the neural network instead of the typical spectrogram to more closely represent the time-frequency representation seen in the cochlea of species such as the echolocating bat \cite{patterson_r_d_efficient_1987, valero_gammatone_2012}.

\begin{figure}
    \centering
    \includegraphics[width=1\linewidth]{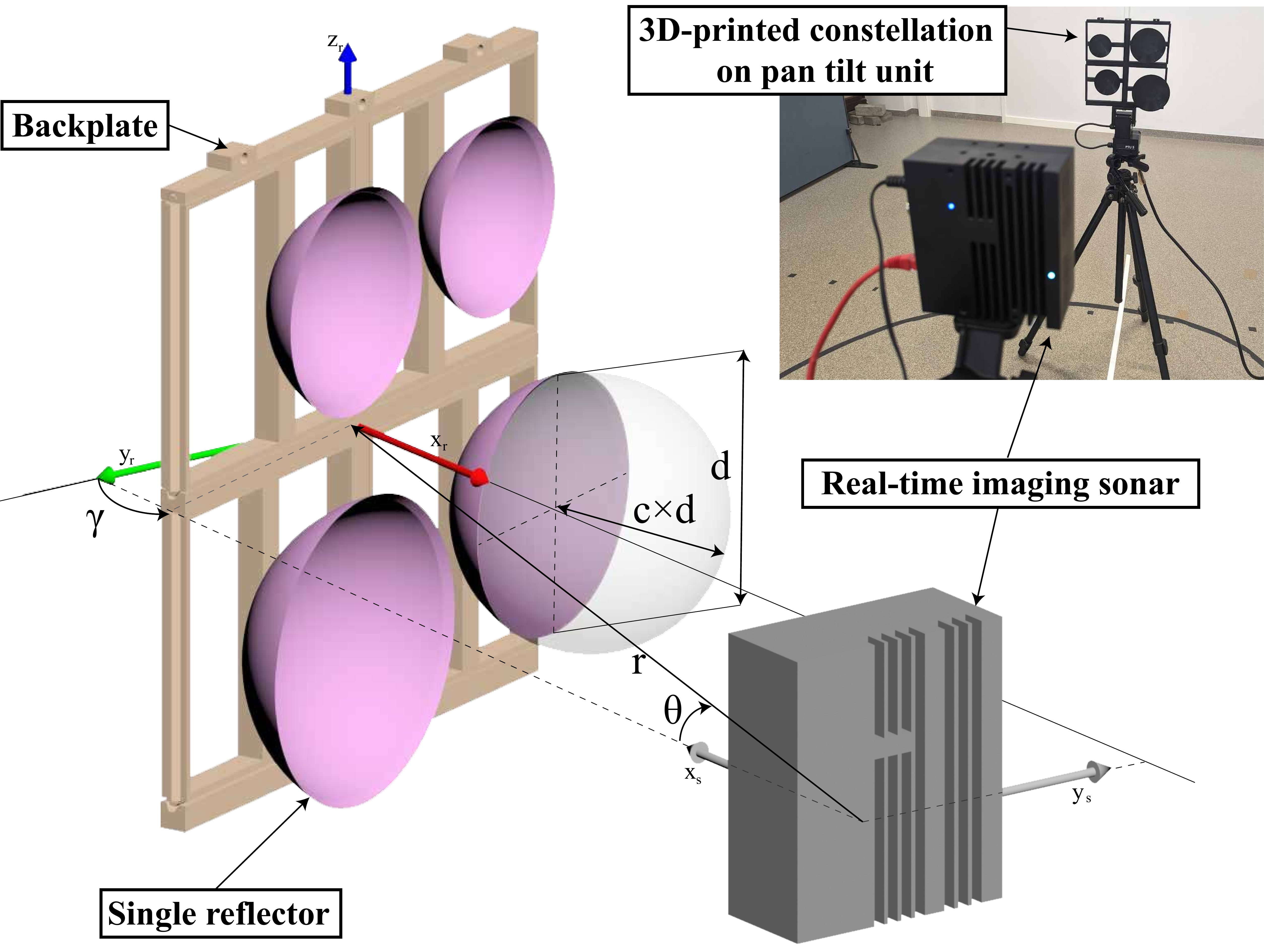}
    \caption{A multi-reflector made of a constellation of four single reflectors with their hollow hemisphere shape extracted from a sphere with radius $d$ cut off by a percentage factor $c$. The eRTIS sensor is always assumed to be parallel to the origin of the landmark. The top right photograph shows the real measurement setup with the pan-tilt device that controls the azimuth and elevation angles of the 3D-printed reflector constellation.}
    \label{fig:multireflector_and_sensor}
\end{figure}

In this paper, we build upon this work to increase the information entropy of acoustic landmarks using deep learning. We move from single reflectors to classifying a constellation of four reflectors, as shown in Figure \ref{fig:multireflector_and_sensor}. This increases the tags' information load fourfold. We implement two techniques to enhance the neural network's accuracy and precision to be able to classify the  constellation of reflectors. Our first contribution is a multi-label CNN classification model designed to simultaneously identify each reflector in the constellation from a single cochleogram. For comparison against single reflector classification results, our second contribution is implementing adaptive beamforming with null-steering to isolate reflectors within a constellation first, allowing for the single-label classification model to classify the constellation in four steps. Finally, we discuss these methods' experimental results and show their benefits and limitations while also providing insights into methodologies that should be explored in the future.

\section{Methodology}\label{sec:methodology}

\subsection{In-Air 3D Sonar \& Semantic Landmarks}\label{subsec:sonarlandmarks}
We use the same sensor as in our previous work for reflector classification \cite{de_backer_detecting_2023, jansen_semantic_2024}: the Embedded Real-Time Imaging Sonar (eRTIS)\cite{kerstens_ertis_2019}. This sensor has a 32-element microphone array in a known, irregularly scattered layout with a single transducer that in this paper emits a broadband FM-sweep of \SI{2.5}{\ms} between \SI{25}{\kilo\hertz} and \SI{80}{\kilo\hertz}. 

The semantic reflectors are based on the original research by Simon et al. \cite{simon_bioinspired_2020}. In this research, they are 3D-printed in ten radii variations, all with tapered edges. The original hollow hemispheres shape with diameter $d$ is cut at a certain percentage factor $c$. In this paper, this is always 66\%. The reflectors were placed in five different constellation configurations with a backplate for easier mounting. One such reflector constellation is illustrated in Figure \ref{fig:multireflector_and_sensor}.

\subsection{Multi-Label Classification}\label{sec:method_multilabel}
Our first contribution is a multi-label CNN classification model designed to simultaneously identify multiple, closely spaced reflector landmarks from a single sonar spectrogram. We follow the same pre-processing steps as discussed in detail in \cite{jansen_semantic_2024}. It is assumed that the reflector constellation is detected, and its orientation $\gamma$ is estimated using the same methodology as described in this previous research, using a separate regression neural network. This system's outcome is a cochleogram steered into the direction of the detected multi-reflector constellation. 

The cochleogram was the input of our CNN in the form of a 40 by 106-pixel image. First, we perform feature extraction using several blocks of convolutional, ReLU, batch normalization, and max-pooling layers. Secondly, several fully connected layers create the classification head and are combined with a sigmoid function to treat each class as an independent binary classification, allowing an input cochleogram to be assigned multiple labels simultaneously. Third, a Binary Cross-Entropy (BCE) layer calculates the aggregated loss. The BCE loss for a single prediction is given by
\begin{equation}
    L_{BCE} = -[t \cdot \log(y) + (1-t) \cdot \log(1-y)]
\end{equation}
where $y$ is the predicted probability from the sigmoid layer and $t$ is the true label (0 or 1). This layer calculates a loss for each class independently and then aggregates these losses. This ensures that the network is penalized for incorrect predictions for each individual label. The ADAM optimization algorithm was used with a learning rate of 0.001, a maximum of 100 epochs was set with validation patience at 20 epochs. Class weights were also used based on their frequency in the training data. 

\subsection{Adaptive Beamforming with Null-Steering}\label{sec:method_nullsteering}
The second method does not attempt to simultaneously identify all reflectors at once but tries to leverage the high accuracy of the single reflector classifier created in \cite{jansen_semantic_2024}. In order to use this single-label classification model, we use adaptive beamforming with null-steering to isolate each reflector within the constellation by placing nulls in the locations of the other three before applying the single reflector classifier. 

To achieve directional sensitivity in our microphone array, we implement a frequency-domain beamforming algorithm. This will find the steering vector $w_d$ to maximize the array's response to the specific direction of interest of the individual reflector being classified. Simultaneously, we have to perform null-steering to minimize its sensitivity in the specified null directions towards the other three reflectors in the constellation \cite{van_trees_optimum_2002}. For each null direction, we compute its corresponding steering vector $C_i$. These vectors form the columns of a constraint matrix $C = [C_0, C_1, C_2]$. We can then find a new weight vector $w$, as close as possible to the desired vector $w_d$ under the constraint that it is orthogonal to the space spanned by the null steering vectors, i.e., $w^HC=0$. To solve this minimization optimization, we are projecting the desired vector onto the null space and subtracting the projection from the original desired vector. First, we compute the projection matrix $P_C$, which projects any vector onto the subspace spanned by the columns of $C$:
\begin{equation}
    P_C = C\left[C^HC\right]^{-1}C^H
\end{equation}
The final, null-steered weight vector $w$, is then calculated by applying the complementary projection:
\begin{equation}
    w^H =  w_d^H(I-P_C)
\end{equation}
This operation effectively removes any components from the desired steering vector that lie in the null directions, thus creating deep nulls in the beam pattern at the specified angles. Doing this, we end up with four distinct cochleograms for each of the reflectors that exist in the constellation, ready to go through the single-label classification model of \cite{jansen_semantic_2024}.

\begin{figure}
    \centering
    \includegraphics[width=1\linewidth]{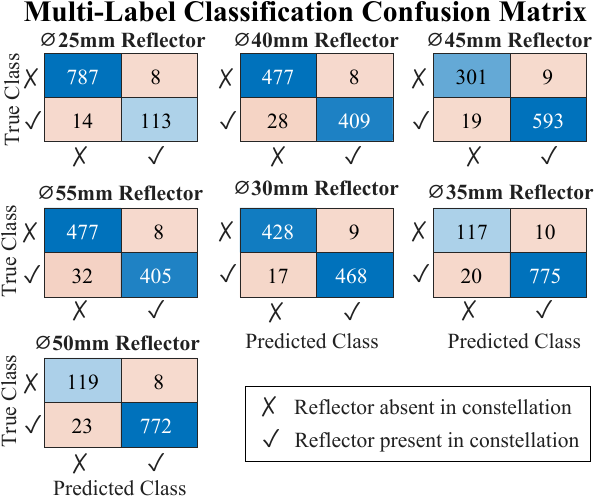}
    \caption{The multi-label classification results for seven different reflector sizes in the form of a confusion matrix. The network reached an overall F1 score of 0.971 and a Jaccard score of 0.928 on the test dataset.}
    \label{fig:classification_confusion}
\end{figure}

\section{Experimental Results}\label{sec:results}
The first experiments were created to analyze the performance of the multi-label classification. These were performed in a cluttered indoor environment with the reflector constellation on a pan-tilt unit mounted on a tripod. The pan-tilt unit varied the landmark's horizontal orientation between \SI{\pm60}{\degree}, and the eRTIS sensor was placed at various positions up to a maximum offset of \SI{3}{\m}. This data collection strategy was designed to ensure the model generalizes well to different viewing angles and distances. Note that the sensor was always on the same axis facing forward towards the tripod of the landmark reflector. Figure \ref{fig:multireflector_and_sensor} illustrates this configuration. After data pre-preprocessing, 21,575 cochleograms were made. The data was split into 64\% training, 16\% validation, and 20\% testing. The predication threshold for positive classification was set to 0.6 after initial experiments indicated this resulted in the highest F1 score and Jaccard index. The network reached an overall F1 score of 0.971 and a Jaccard score of 0.928 on the test dataset, demonstrating a high degree of accuracy in identifying the constellation's components. The confusion matrix of this result is shown in Figure \ref{fig:classification_confusion}.

\begin{figure}
    \centering
    \includegraphics[width=1\linewidth]{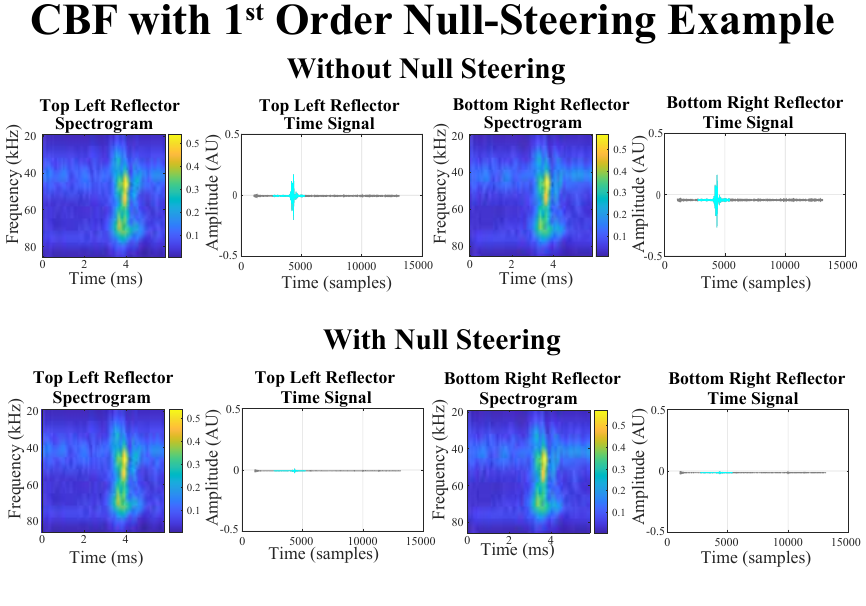}
    \caption{An example result of the adaptive-beamforming with null steering on a measurement where a constellation reflector was present. Showing both the time and frequency domain results for the top left and bottom right reflectors respectively with and without null steering applied.}
    \label{fig:nullsteering}
\end{figure}

For comparison, we explored adaptive beamforming with null-steering to isolate reflectors within a constellation, theoretically allowing for single-label classification models to classify the constellation in multiple steps. However, our initial results suggest challenges in achieving sufficient isolation in practice. An ablation study where we compared the adaptive-beamforming with and without the null steering showed no remarkable difference in the resulting data as seen in the example in Figure \ref{fig:nullsteering}. Using the resulting cochleograms in our single-label classifier resulted in an overall accuracy below 45\%, indicating that this approach was not effective with our current implementation.

\section{Discussion \& Conclusions}\label{sec:conclusions}
In this work, we investigated the feasibility of classifying constellations of passive ultrasonic reflectors to move towards in-air acoustic tags with a higher information capacity. Our results demonstrate that a multi-label CNN classifier can perform this task with reasonable success on our limited test dataset, correctly identifying the presence of multiple reflectors simultaneously from a single sonar measurement. This finding is a promising first step, validating the core concept of decoding complex, overlapping acoustic echoes. However, our exploration of adaptive beamforming with null-steering to isolate individual reflectors for a single-label classifier, yielded less effective results. The initial tests suggest significant challenges in achieving sufficient acoustic isolation in practice. This indicates that the reflectors' physical design and relative spacing within a constellation are critical factors that heavily influence echo separation and, consequently, single-label classification performance. The reflectors' close proximity likely creates complex constructive and destructive interference patterns that are challenging to disentangle with first-order null-steering, highlighting a key area for future investigation. 

Based on our findings and further literature review, several promising avenues exist for future work to increase the information entropy of ultrasonic passive reflector constellations. A crucial area of research is the design of the reflectors themselves. To achieve greater differentiation in the resulting echoes, future studies should explore more significant variations in reflector geometry beyond simple radii, investigate the impact of different surface materials on acoustic reflectance, and delve into the emerging field of acoustic meta-materials. Such materials could be engineered to produce highly unique and easily distinguishable echo signatures, dramatically expanding the potential encoding capacity of a landmark.

\newpage

On the classification front, while our initial deep learning model seemed promising, its performance could be substantially improved and be more robust in larger, more varied real-world scenarios. Data augmentation techniques are known to enhance model robustness and generalization \cite{eisele_convolutional_2023}. We have developed a preliminary MATLAB toolbox for this purpose, but time constraints prevented a thorough investigation into its impact on our results \cite{jansen_spectrogram_nodate}. Furthermore, future work should explore different input representations beyond the time-frequency domain. Using the raw time-domain echo signals directly for classification with architectures like 1D-CNNs, could capture subtle temporal features lost in the transformation to a spectrogram. Moreover, leveraging complex-valued neural network architectures\cite{hirose_complex-valued_2012} could allow the model to use the rich phase information present in the sonar signal, which is currently discarded when only amplitude is considered \cite{eisele_relevance_2024}. Finally, to build upon our initial null-steering experiments, more advanced adaptive beamforming techniques could be implemented. Methods such as Linearly Constrained Minimum Variance (LCMV)\cite{markovich_multichannel_2009} beamforming are designed to place deeper, more precise nulls and could prove more effective at isolating the echoes from individual reflectors within a tightly packed constellation \cite{souden_study_2010}.

\bibliographystyle{IEEEtran}
\bibliography{main.bib}

\begin{thebibliography}{10}
\providecommand{\url}[1]{#1}
\csname url@samestyle\endcsname
\providecommand{\newblock}{\relax}
\providecommand{\bibinfo}[2]{#2}
\providecommand{\BIBentrySTDinterwordspacing}{\spaceskip=0pt\relax}
\providecommand{\BIBentryALTinterwordstretchfactor}{4}
\providecommand{\BIBentryALTinterwordspacing}{\spaceskip=\fontdimen2\font plus
\BIBentryALTinterwordstretchfactor\fontdimen3\font minus
  \fontdimen4\font\relax}
\providecommand{\BIBforeignlanguage}[2]{{%
\expandafter\ifx\csname l@#1\endcsname\relax
\typeout{** WARNING: IEEEtran.bst: No hyphenation pattern has been}%
\typeout{** loaded for the language `#1'. Using the pattern for}%
\typeout{** the default language instead.}%
\else
\language=\csname l@#1\endcsname
\fi
#2}}
\providecommand{\BIBdecl}{\relax}
\BIBdecl

\bibitem{thrun_probabilistic_2005}
S.~Thrun, W.~Burgard, and D.~Fox, \emph{Probabilistic robotics}, ser.
  Intelligent robotics and autonomous agents.\hskip 1em plus 0.5em minus
  0.4em\relax Cambridge, Mass: MIT Press, 2005, oCLC: ocm58451645.

\bibitem{grisetti_tutorial_2010}
\BIBentryALTinterwordspacing
G.~Grisetti, R.~Kümmerle, C.~Stachniss, and W.~Burgard, ``A {Tutorial} on
  {Graph}-{Based} {SLAM},'' \emph{IEEE Intelligent Transportation Systems
  Magazine}, vol.~2, no.~4, pp. 31--43, 2010. [Online]. Available:
  \url{https://ieeexplore.ieee.org/document/5681215}
\BIBentrySTDinterwordspacing

\bibitem{yoneda_automated_2019}
\BIBentryALTinterwordspacing
K.~Yoneda, N.~Suganuma, R.~Yanase, and M.~Aldibaja, ``Automated driving
  recognition technologies for adverse weather conditions,'' \emph{IATSS
  Research}, vol.~43, no.~4, pp. 253--262, Dec. 2019. [Online]. Available:
  \url{https://www.sciencedirect.com/science/article/pii/S0386111219301463}
\BIBentrySTDinterwordspacing

\bibitem{zhang_perception_2023}
\BIBentryALTinterwordspacing
Y.~Zhang, A.~Carballo, H.~Yang, and K.~Takeda, ``Perception and sensing for
  autonomous vehicles under adverse weather conditions: {A} survey,''
  \emph{ISPRS Journal of Photogrammetry and Remote Sensing}, vol. 196, pp.
  146--177, Feb. 2023. [Online]. Available:
  \url{https://www.sciencedirect.com/science/article/pii/S0924271622003367}
\BIBentrySTDinterwordspacing

\bibitem{everett_sensors_1995}
H.~R. Everett, \emph{Sensors for {Mobile} {Robots}: {Theory} and
  {Application}}.\hskip 1em plus 0.5em minus 0.4em\relax A K Peter, CRC Press,
  1995.

\bibitem{jansen_real-time_2022}
\BIBentryALTinterwordspacing
W.~Jansen, D.~Laurijssen, and J.~Steckel, ``\BIBforeignlanguage{en}{Real-{Time}
  {Sonar} {Fusion} for {Layered} {Navigation} {Controller}},''
  \emph{\BIBforeignlanguage{en}{Sensors}}, vol.~22, no.~9, p. 3109, Jan. 2022.
  [Online]. Available: \url{https://www.mdpi.com/1424-8220/22/9/3109}
\BIBentrySTDinterwordspacing

\bibitem{laurijssen_hiris_2024}
\BIBentryALTinterwordspacing
D.~Laurijssen, W.~Daems, and J.~Steckel, ``{HiRIS}: {An} {Airborne} {Sonar}
  {Sensor} {With} a 1024 {Channel} {Microphone} {Array} for {In}-{Air}
  {Acoustic} {Imaging},'' \emph{IEEE Access}, vol.~12, pp. 51\,786--51\,795,
  2024. [Online]. Available:
  \url{https://ieeexplore.ieee.org/document/10491247}
\BIBentrySTDinterwordspacing

\bibitem{allevato_air-coupled_2022}
\BIBentryALTinterwordspacing
G.~Allevato, M.~Rutsch, J.~Hinrichs, C.~Haugwitz, R.~Müller, M.~Pesavento, and
  M.~Kupnik, ``Air-{Coupled} {Ultrasonic} {Spiral} {Phased} {Array} for
  {High}-{Precision} {Beamforming} and {Imaging} {\textbar} {IEEE} {Journals}
  \& {Magazine} {\textbar} {IEEE} {Xplore},'' \emph{IEEE Open Journal of
  Ultrasonics, Ferroelectrics, and Frequency Control}, vol.~2, pp. 40--54, Jan.
  2022. [Online]. Available: \url{https://ieeexplore.ieee.org/document/9678369}
\BIBentrySTDinterwordspacing

\bibitem{kerstens_ertis_2019}
\BIBentryALTinterwordspacing
R.~Kerstens, D.~Laurijssen, and J.~Steckel, ``{eRTIS}: {A} {Fully} {Embedded}
  {Real} {Time} {3D} {Imaging} {Sonar} {Sensor} for {Robotic} {Applications},''
  in \emph{2019 {International} {Conference} on {Robotics} and {Automation}
  ({ICRA})}.\hskip 1em plus 0.5em minus 0.4em\relax Montreal, QC, Canada: IEEE,
  May 2019, pp. 1438--1443. [Online]. Available:
  \url{https://ieeexplore.ieee.org/document/8794419/}
\BIBentrySTDinterwordspacing

\bibitem{eisele_relevance_2024}
\BIBentryALTinterwordspacing
J.~Eisele, A.~Gerlach, M.~Maeder, and S.~Marburg, ``Relevance of phase
  information for object classification in automotive ultrasonic sensing using
  convolutional neural networks,'' \emph{The Journal of the Acoustical Society
  of America}, vol. 155, no.~2, pp. 1060--1070, Feb. 2024. [Online]. Available:
  \url{https://doi.org/10.1121/10.0024753}
\BIBentrySTDinterwordspacing

\bibitem{werner_g_neubauer_acoustic_1986}
{Werner G. Neubauer}, \emph{Acoustic {Reflection} from {Surfaces} and
  {Shapes}}.\hskip 1em plus 0.5em minus 0.4em\relax Naval Research Lab
  Washington DC, Jun. 1986.

\bibitem{simmons_acoustic_1989}
J.~A. Simmons and L.~Chen, ``The acoustic basis for target discrimination by
  {FM} echolocating bats,'' \emph{The Journal of the Acoustical Society of
  America}, vol.~86, no.~4, pp. 1333--1350, Oct. 1989.

\bibitem{dror_three-dimensional_1995}
\BIBentryALTinterwordspacing
I.~E. Dror, M.~Zagaeski, and C.~F. Moss, ``Three-{Dimensional} target
  recognition via sonar: {A} neural network model,'' \emph{Neural Networks},
  vol.~8, no.~1, pp. 149--160, Jan. 1995. [Online]. Available:
  \url{https://www.sciencedirect.com/science/article/pii/089360809400057S}
\BIBentrySTDinterwordspacing

\bibitem{kroh_classification_2019}
\BIBentryALTinterwordspacing
P.~K. Kroh, R.~Simon, and S.~J. Rupitsch,
  ``\BIBforeignlanguage{en}{Classification of {Sonar} {Targets} in {Air}: {A}
  {Neural} {Network} {Approach}},'' \emph{\BIBforeignlanguage{en}{Sensors}},
  vol.~19, no.~5, p. 1176, Jan. 2019. [Online]. Available:
  \url{https://www.mdpi.com/1424-8220/19/5/1176}
\BIBentrySTDinterwordspacing

\bibitem{ayrulu_neural_2001}
\BIBentryALTinterwordspacing
B.~Ayrulu and B.~Barshan, ``Neural networks for improved target differentiation
  and localization with sonar,'' \emph{Neural Networks}, vol.~14, no.~3, pp.
  355--373, Apr. 2001. [Online]. Available:
  \url{https://www.sciencedirect.com/science/article/pii/S089360800100017X}
\BIBentrySTDinterwordspacing

\bibitem{simon_bioinspired_2020}
\BIBentryALTinterwordspacing
R.~Simon, S.~Rupitsch, M.~Baumann, H.~Wu, H.~Peremans, and J.~Steckel,
  ``Bioinspired sonar reflectors as guiding beacons for autonomous
  navigation,'' \emph{Proceedings of the National Academy of Sciences}, vol.
  117, no.~3, pp. 1367--1374, Jan. 2020. [Online]. Available:
  \url{https://www.pnas.org/doi/full/10.1073/pnas.1909890117}
\BIBentrySTDinterwordspacing

\bibitem{de_backer_detecting_2023}
\BIBentryALTinterwordspacing
M.~de~Backer, W.~Jansen, D.~Laurijssen, R.~Simon, W.~Daems, and J.~Steckel,
  ``Detecting and {Classifying} {Bio}-{Inspired} {Artificial} {Landmarks}
  {Using} {In}-{Air} {3D} {Sonar},'' in \emph{2023 {IEEE} {SENSORS}}, Oct.
  2023, pp. 1--4, iSSN: 2168-9229. [Online]. Available:
  \url{https://ieeexplore.ieee.org/document/10325158}
\BIBentrySTDinterwordspacing

\bibitem{jansen_semantic_2024}
\BIBentryALTinterwordspacing
W.~Jansen and J.~Steckel, ``Semantic {Landmark} {Detection} \& {Classification}
  {Using} {Neural} {Networks} {For} {3D} {In}-{Air} {Sonar},'' in \emph{2024
  {IEEE} {SENSORS}}.\hskip 1em plus 0.5em minus 0.4em\relax Kobe, Japan: IEEE,
  Oct. 2024, pp. 1--4. [Online]. Available:
  \url{https://ieeexplore.ieee.org/document/10785063/}
\BIBentrySTDinterwordspacing

\bibitem{patterson_r_d_efficient_1987}
{Patterson, R. D.}, {Nimmo-Smith, I.}, {Holdsworth, J.}, and {Rice, P.}, ``An
  efficient auditory filterbank based on the gammatone function,'' \emph{a
  meeting of the IOC Speech Group on Auditory Modelling at RSRE}, vol.~2,
  no.~7, Dec. 1987.

\bibitem{valero_gammatone_2012}
\BIBentryALTinterwordspacing
X.~Valero and F.~Alias, ``Gammatone {Cepstral} {Coefficients}: {Biologically}
  {Inspired} {Features} for {Non}-{Speech} {Audio} {Classification},''
  \emph{IEEE Transactions on Multimedia}, vol.~14, no.~6, pp. 1684--1689, Dec.
  2012. [Online]. Available:
  \url{https://ieeexplore.ieee.org/abstract/document/6202347}
\BIBentrySTDinterwordspacing

\bibitem{van_trees_optimum_2002}
\BIBentryALTinterwordspacing
H.~L. Van~Trees, \emph{\BIBforeignlanguage{en}{Optimum {Array} {Processing}:
  {Part} {IV} of {Detection}, {Estimation}, and {Modulation} {Theory}}},
  1st~ed.\hskip 1em plus 0.5em minus 0.4em\relax Wiley, Mar. 2002. [Online].
  Available: \url{https://onlinelibrary.wiley.com/doi/book/10.1002/0471221104}
\BIBentrySTDinterwordspacing

\bibitem{eisele_convolutional_2023}
\BIBentryALTinterwordspacing
J.~Eisele, A.~Gerlach, M.~Maeder, and S.~Marburg,
  ``\BIBforeignlanguage{en}{Convolutional neural network with data augmentation
  for object classification in automotive ultrasonic sensing},''
  \emph{\BIBforeignlanguage{en}{The Journal of the Acoustical Society of
  America}}, vol. 153, no.~4, p. 2447, Apr. 2023. [Online]. Available:
  \url{https://pubs.aip.org/jasa/article/153/4/2447/2884903/Convolutional-neural-network-with-data}
\BIBentrySTDinterwordspacing

\bibitem{jansen_spectrogram_nodate}
\BIBentryALTinterwordspacing
W.~Jansen, ``Spectrogram {Augmenter} {Matlab} {Toolbox}.'' [Online]. Available:
  \url{https://nl.mathworks.com/matlabcentral/fileexchange/172765-spectrogram-augmenter}
\BIBentrySTDinterwordspacing

\bibitem{hirose_complex-valued_2012}
\BIBentryALTinterwordspacing
A.~Hirose, \emph{\BIBforeignlanguage{en}{Complex-{Valued} {Neural}
  {Networks}}}, ser. Studies in {Computational} {Intelligence}.\hskip 1em plus
  0.5em minus 0.4em\relax Berlin, Heidelberg: Springer Berlin Heidelberg, 2012,
  vol. 400. [Online]. Available:
  \url{https://link.springer.com/10.1007/978-3-642-27632-3}
\BIBentrySTDinterwordspacing

\bibitem{markovich_multichannel_2009}
\BIBentryALTinterwordspacing
S.~Markovich, S.~Gannot, and I.~Cohen, ``Multichannel {Eigenspace}
  {Beamforming} in a {Reverberant} {Noisy} {Environment} {With} {Multiple}
  {Interfering} {Speech} {Signals},'' \emph{IEEE Transactions on Audio, Speech,
  and Language Processing}, vol.~17, no.~6, pp. 1071--1086, Aug. 2009.
  [Online]. Available: \url{http://ieeexplore.ieee.org/document/5109760/}
\BIBentrySTDinterwordspacing

\bibitem{souden_study_2010}
\BIBentryALTinterwordspacing
M.~Souden, J.~Benesty, and S.~Affes, ``A {Study} of the {LCMV} and {MVDR}
  {Noise} {Reduction} {Filters},'' \emph{IEEE Transactions on Signal
  Processing}, vol.~58, no.~9, pp. 4925--4935, Sep. 2010. [Online]. Available:
  \url{http://ieeexplore.ieee.org/document/5482097/}
\BIBentrySTDinterwordspacing

\end{thebibliography}

\end{document}